\documentclass[12pt]{iopart}
\usepackage{color}
\usepackage{citesort}
\usepackage{iopams} 
\usepackage{subfigure}
\usepackage{graphicx}
\usepackage{epstopdf}
\expandafter\let\csname equation*\endcsname\relax
\expandafter\let\csname endequation*\endcsname\relax
\usepackage{amsmath,amssymb}    
\usepackage[numbers]{natbib}
\usepackage[titletoc,toc,title]{appendix}
\usepackage{graphicx}
\usepackage{subfig}
\usepackage{caption}
\usepackage{sidecap}
\usepackage[applemac]{inputenc}
\usepackage{natbib}
\usepackage{graphicx}           
\usepackage{flafter}            
\usepackage{amsmath,amssymb}    
\usepackage{bm}                 
\usepackage{memhfixc}           
\usepackage[activate]{pdfcprot} 
\usepackage{pdfsync}            

\usepackage{verbatim} 

\newtheorem{lemma}{Lemma}
\newtheorem{proof}{Proof}

\begin{document}
\title[Solving TAP Equations for Ising Models with General Invariant Random Matrices]{A Theory of Solving TAP Equations for Ising Models with General Invariant Random Matrices}
\vspace{0.1pc}
\author{Manfred Opper$^1$, Burak \c{C}akmak$^2$ and Ole Winther$^3$}
\vspace{0.3pc}
\address{$^1$ Department of Artificial Intelligence, Technische Universit\"{a}t Berlin, Marchstraße 23, $~~$Berlin 10587, Germany}
\vspace{0.15pc}
\address{$^2$ Department of Electronic Systems, Aalborg Universitet, Fredrik Bajers Vej 7, \\ $~~$Alborg 9220, Denmark}
\vspace{0.15pc}
\address{$^3$ DTU Compute, Danmarks Tekniske Universitet, Anker Engelunds Vej 1, \\ $~~$Lyngby 2800, Denmark}
\vspace{0.25pc}
\eads{\mailto{manfred.opper@tu-berlin.de}, \mailto{buc@es.aau.dk} and \mailto{olwi@dtu.dk}}

\begin{abstract}
We consider the problem of solving TAP mean field equations by iteration for Ising models
with coupling matrices that are drawn at random from general invariant ensembles. We develop an analysis of iterative algorithms using a dynamical functional approach that in the thermodynamic limit yields an effective dynamics of a single variable trajectory. Our main novel contribution is the expression for the implicit memory term of the dynamics for general invariant ensembles. By subtracting these terms, that depend on magnetizations at previous time steps, the implicit memory terms cancel making the iteration dependent on a Gaussian distributed field only. The TAP magnetizations are stable fixed points if an AT stability criterion is fulfilled. We illustrate our method explicitly for coupling matrices drawn from the random orthogonal ensemble.
\end{abstract}

\vspace{1pc}




\bibliographystyle{plain}
\def\mathlette#1#2{{\mathchoice{\mbox{#1$\displaystyle #2$}}%
                               {\mbox{#1$\textstyle #2$}}%
                               {\mbox{#1$\scriptstyle #2$}}%
                               {\mbox{#1$\scriptscriptstyle #2$}}}}
\newcommand{\matr}[1]{\mathlette{\boldmath}{#1}}
\newcommand{\RR}{\mathbb{R}}
\newcommand{\CC}{\mathbb{C}}
\newcommand{\NN}{\mathbb{N}}
\newcommand{\ZZ}{\mathbb{Z}}
\newcommand{\at}[2][]{#1|_{#2}}
\vspace{2pc}

\section{Introduction}
TAP equations provide generalized mean field equations for statistical physics models with random, infinite range interactions which (under certain conditions) are assumed to be exact in the limit of an infinite system \cite{Mezard}. In recent years there has been an increasing interest in such equations within and also outside of the statistical physics community. This is partly due to the fact that the TAP approach can be applied to statistical inference in probabilistic models in information theory \cite{Kabashima}, \cite{Bayati}, statistics \cite{adatap} and machine learning \cite{OW0}, \cite{Minka1}. Originally developed by D.J. Thouless, P. Anderson and R. Palmer for the Sherrington Kirkpatrick (SK) model of an Ising spin--glass \cite{TAP}, the TAP approach has been generalised to a variety of other problems \cite{adatap}. This includes models with continuous variables rather than Ising spins but also cases where the independent random interactions are replaced by other, more structured
statistical ensembles that allow for certain dependencies. 

While methods for deriving the TAP approach for different models are now well established it is not necessarily clear how the resulting system of nonlinear equations can be solved efficiently. A naive algorithm based on a simple iteration of the equations usually fails to achieve convergence.  This problem has been addressed by a paper of Bolthausen for the case of the SK model \cite{Bolthausen}. He has analyzed the dynamics of iterations rigorously and shown how the iterations can be altered in order to achieve exponential convergence (above the so--called AT line of stability). Other ideas to arrive at a convergent method are based on taking the limit of dense couplings in belief propagation algorithms, see \cite{Kabashima}, \cite{Donoha}, and \cite{Florent}, (for rigorous analyses \cite{Bayati} and \cite{bayati15}). Unfortunately, for this approach it is necessary to augment the original variables by auxiliary ones, such that the interactions in the new model are independent. For example, the Hopfield model can be represented by a bipartite graph of Ising spins and continuous variables. It is not clear how such a method should be set up for a matrix of interactions with more general statistical dependencies.  Also, taking the dense coupling limit of an approximate message passing (AMP) algorithm valid for sparse coupling problems will not always lead to the correct dense coupling algorithm. Here we will construct a theory for dynamics using dense couplings as the starting point.

In this paper we will address the problem of solving TAP equations for Ising models
with dense random coupling matrices with a general invariant probability distribution. 
Our analysis is based on dynamical mean field theory which allows to study the dynamics of
iterative algorithms in the thermodynamic limit by a suitable average over the ensemble of couplings. 
It turns out that by including certain memory terms in the iteration, the effective field in the dynamics 
becomes a simple Gaussian random variable suggesting that the dynamics might converge.
The explicit form of the memory terms depends explicitly on the statistical ensemble of couplings.
We show that our method reproduces previous convergent algorithms for SK and Hopfield models.
We also work out the details of our theory for the spin model with orthogonal random couplings \cite{Parisi, Enzo}.
Simulations of the resulting algorithms show exponential convergence above a line of stability which
can be identified with the so--called AT line. 

The paper is organized as follows: in Section II we introduce the general random matrix formulation for the TAP equations. In Section III we present the results of dynamical functional theory. In Section IV, we introduce ``the single-step memory construction" iterative algorithm for solving TAP equations. Section V is devoted to the derivation of the AT stability condition. Discussions and outlooks are presented in Section VI. Lengthy technical derivations are deferred to the Appendix.

\section{General Invariant Random Matrix Ensembles} \label{GIRME}
We will consider Ising models with pairwise interactions given by the Gibbs distribution 
for the spins $\matr S = (S_1, \ldots, S_N)$
\begin{equation}
P(\matr S) = \frac{1}{Z} \exp\left[\sum_{i < j}^N J_{ij} S_i S_j +  \sum_i^N h_i S_i\right]
\label{Gibbs_Ising}
\end{equation}
with $Z$ denoting the normalization constant. We are interested in the case, where the matrix $\matr {J}$ is random with the condition that the marginal density of couplings $p(J_{ij})$ is the same for all pairs $(i,j)$ but couplings might be dependent random variables. A simple way for defining such class of random matrices is via the so--called {\em invariant} ensembles \cite{Percy}. A random matrix $\matr {\tilde J}$ is called invariant if it has the same probability distribution as $\matr V^\dagger\matr {\tilde J}\matr V$ for an orthogonal matrix $\matr V$ which is independent of $\matr {\tilde J}$ \footnotetext{Here $(\cdot)^\dagger$ denotes transposition.}. Equivalently, it admits the spectral decomposition 
\begin{equation}
\matr {\tilde J}= \matr O^\dagger \matr \Lambda \matr O\label{spectral}
\end{equation}
where $\matr O$ is Haar distributed (i.e.\ it is a random orthogonal matrix) and independent of the diagonal matrix $\matr \Lambda$. This characterization of invariant matrices involves the diagonal elements $\tilde{J}_{ii}$ which is absent in (\ref{Gibbs_Ising}). However, one can show that as $N$ tends to infinity if the spectrum of $\matr {\tilde J}$ converges almost surely to a compactly supported distribution such that the smallest and largest eigenvalue of $\matr {\tilde J}$ converge almost surely to the infimum and supremum of the support, respectively, we have (in the almost sure sense) 
$
{\tilde J}_{ii}-\frac{1}{N}{\tr}(\matr {\tilde J})\to 0, \forall i$ as $N$ tends to infinity, see \ref{Profof2}. In other words the diagonal elements of an invariant matrix converge to the same deterministic limit. Thus, we may define asymptotically invariant couplings $\matr J$ as $J_{ij} =  \beta\tilde J_{ij}$ for $i\neq j$ and $J_{ii}=0$, where we include an inverse temperature factor $\beta$ in the definition. The couplings for the standard SK and Hopfield models belong to this class of matrices for $(-\matr{\tilde J})$ being the {Gaussian} Wigner matrix and the {null Wishart} matrix {(i.e. a sample covariance matrix of independent Gaussian random vectors whose entries are zero mean, independent and identical distributed)}, respectively. 
\subsection{The Generating Function and the R-transform} \label{Rtrasform} 
We will later need the generating function of asymptotically invariant random matrices given by \cite{Enzo,Guionnet}
\begin{equation}
{\rm G}(\matr Q)\triangleq \lim_{N\to \infty}\frac{1}{N} \log \left<e^{\frac{N}{2}\mbox{tr} (\bm{Q}\bm{J} )}\right>_{\matr J} \label{G1}
\end{equation}
with symmetric matrix $\matr Q$ having a finite rank. Since we believe that our paper might be of interest
to researchers with an information theory background, we will briefly mention how  ${\rm G}$ is related
to quantities which are well known in the theory of free probability \cite{Voi92}, which is a powerful approach to random matrix theory. 
Setting 
\begin{equation}
{\rm G}(x)=\frac{1}{2}\int_{0}^{x}{\rm d}\omega\; {\rm R}(\omega) \label{G2}
\end{equation}
one can show that $\rm R$ equals the so--called R-transform (in the theory of free probability) of the limiting spectrum.  
Its formal definition can be given in terms of the Cauchy transform: let ${\rm P}$ denote the limiting spectrum of $\matr J$. Moreover let
\begin{equation}
{\rm M}(z)\triangleq\frac{1}{z}+\sum_{n=1}^{\infty}\frac{m_n}{z^{n+1}} \ , 
\end{equation}
where $m_n$ is $n$th order moments of the limiting spectrum $\rm P$, i.e.\  $m_n= \int {\rm dP}(\lambda)\; \lambda^n$. Then the R-transform of $\rm P$ is given by
\begin{equation}
{\rm R}(x)= {\rm M}^{-1}(x)-\frac{1}{x}
\end{equation}
with ${\rm M}^{-1}$ denoting the composition inverse of $\rm M$. It admits the power series expansion
\begin{equation}
{\rm R}(x)=\sum_{n=1}^{\infty}c_nx^{n-1} \label{G3} \ , 
\end{equation}
where $c_n$ are known as the \emph{free cumulants} of $\rm P$. For example, the first two free cumulants $c_1$ and $c_2$ are the mean and variance of the distribution $\rm P$, respectively, i.e. \ $c_1=m_1$ and $c_2=m_2-m_1^2$. For details we refer the reader to \cite{Hiai}.

In the sequel we give the R-transforms for some random matrix ensembles that we will discuss later.  We first point out the simple identity
\begin{equation}
{\rm R }(x)= \beta({\rm \tilde R}(\beta x)-\tilde c_1) \ . \label{conv}
\end{equation}
In this expression ${\rm \tilde R}$ denotes the R-transform of the limiting
spectrum of $\matr {\tilde J}$ and $\tilde c_1={\rm \tilde R}(0)$. We next provide the explicit form of ${\rm R}(x)$ for the SK, Hopfield and random orthogonal \cite{Parisi, Enzo} models, respectively: i) For the SK case we have a (symmetric) matrix $\matr {\tilde J}$ whose diagonal entries are zero and the upper-triangle entries are independent identically distributed (iid) Gaussian with zero mean and variance $1/N$. Note that in this case we have $\matr J=\beta \matr{\tilde J}$; so that ${\rm R}(x)=\beta^2 x$ \cite{Hiai}; ii) 
For the Hopfield model, we consider entries of an $(N/\alpha)\times N$ matrix $\matr H$ that are iid and Gaussian with zero mean and variance $\alpha/N$. Moreover let $\matr {\tilde J}=-\matr H^\dagger \matr H$, i.e.\  $(-\matr{\tilde J})$ is a {null} Wishart matrix whose limiting spectrum is given by the Mar\u{c}enko-Pastur distribution. By invoking the R-transform of the Mar\u{c}enko-Pastur distribution, see e.g.~\cite{Hiai}, we have 
\begin{equation}
{\rm R}(x)= \frac{\beta^2 \alpha x}{1+\beta \alpha x}; \label{hopfield}
\end{equation}
iii) Finally, for the random orthogonal case
we consider a spectral decomposition \eqref{spectral} such that $\matr {\tilde J}= \matr O^\dagger \matr \Lambda\matr O$. Here the diagonal entries of the diagonal matrix $\matr \Lambda$ are composed of $\pm 1$ such that ${\rm tr}(\matr \Lambda)=0$. Then, one can easily show that
\begin{equation}
{\rm R}(x)= \frac{-1+\sqrt{1+4\beta^2x^2}}{2x} \ \label{ROM}.
\end{equation}
The latter result was given in \cite{Parisi}.

\subsection{TAP Equations for General Invariant Couplings}
TAP equations are a set of self-consistent equations for the vector of magnetisations $\matr m = \langle \matr S \rangle$ 
where the brackets denote expectation w.r.t. the Gibbs distribution (\ref{Gibbs_Ising}). For a 
general invariant ensemble they have been derived first in \cite{Parisi} using the large $N$ scaling of a perturbation 
expansion.  A second derivation using the cavity method and the large $N$ limit of the `adaptive' TAP equations can be found \cite{adatap}. We have  provided a more rigorous derivation of the transition from ``adaptive TAP"
to the self-averaging limit using random matrix theory in Appendix~B. 
The resulting TAP equations read
\begin{align}
\matr m &=\tanh (\matr \psi) \label{Tap1}\\
\matr\psi &=\matr h+\matr J \matr m-{\rm R}(1-q)\matr m \;  \label{Tap2} 
\end{align}
where $q\triangleq\frac{1}{N}\matr m^\dagger \matr m$ and $\matr h$ is the vector of nonrandom external fields. 
Note, that the only dependency on the random matrix ensemble is via the R--transform 
${\rm R}(1-q)$ in the so--called Onsager term which is a correction to the naive mean field term $\matr J \matr m$.
One can show that $\Psi_i$
is the mean of the cavity field. Furthermore, following the calculations of \cite{adatap} 
one finds that $\Psi_i$ is Gaussian distributed (in the large $N$ limit) with respect to the random couplings $\bm{J}$ with mean $h_i$ and variance 
\begin{equation}
\left<(\Psi_i-h_i)^2\right>=q{\rm R}'(1-q). \label{rs}
\end{equation}
Hence, the subtraction of the Onsager term ${\rm R}(1-q) \matr m$ from the mean field $\matr J \matr m$ makes the remainder Gaussian. We will next transfer the idea of a Gaussian field from the static solutions to the dynamics of an algorithm.

\section{The Results of Dynamical Functional Theory}
Dynamical properties of disordered systems can be computed by the method of dynamical functionals \cite{Martin}. In the limit $N\to\infty$ this method provides us with exact results for the marginal distribution of 
a trajectory of a single variable (in our case a magnetization $m_i(t)$), when we define a dynamics
of an algorithm for the solution of the TAP equations. As a  typical result of such a calculation one finds that the `field'
$\matr J \matr m(t)$ becomes a sum of a Gaussian term and a memory term which includes the magnetizations at
all previous times. This memory often makes the dynamics of disordered systems highly complex allowing e.g.\ for a
persistent dependency on the initial conditions and thus a failure to converge to a unique fixed point. Hence, we propose to
introduce {\em explicit} memory terms which are chosen to cancel the {\em implicit} memory terms 
derived from the dynamical functional theory. In such a way, at each time step, the update of the magnetization
for the algorithm involve a Gaussian distributed random field only and we expect that we might obtain 
good convergence results. This Gaussian property of the effective dynamical field was already shown for a Hopfield model in \cite{Kabashima} and \cite{Donoha} (and proved in \cite{Bayati}) and reappeared in Bolthausen's iterative construction of solutions to the TAP equations for the SK model in \cite{Bolthausen}.

We start with defining a set of dynamical equations which could serve as a candidate algorithm 
for solving the TAP equations for invariant random coupling matrices
\begin{align}
\matr m(t) &= f_t\left(\{\matr \gamma(\tau),\matr m(\tau)\}_{\tau=0}^{t-1}\right) \label{d1} \\
\matr \gamma(t) &= \matr {h}+\matr J \matr m(t)\label{d2}
\end{align} 
for $\tau = 0,\ldots, t$ which depend on the field $\matr J \matr m(t)$ and the previous local magnetizations $\matr m(\tau)$. Here $f_t$ is an appropriate sequence of non-linear scalar functions. Our goal is to get the statistics of a single trajectory of \eqref{d1}-\eqref{d2}, when $\matr J$ is a random matrix with generating function (\ref{G1}). To do so we make use of the dynamical functional theory (DFT) analysis as described in \cite{Eisfeller,Henkel} which is a discrete time version of the method of \cite{Martin}. We also refer the reader to \cite{Mimura} where DFT was used to analyze the AMP algorithm in the context of the CDMA communication algorithm.  

We introduce the generating functional corresponding to the dynamics \eqref{d1}--\eqref{d2} as 
\begin{align}
Z(\{\matr l(t)\})=&\int \prod_{t=0}^{T-1}\left\{ {\rm d}\matr m(t){\rm d}\matr \gamma(t)\;
\delta (\matr m(t)- f_t\left(\{\matr\gamma(\tau), \matr m(\tau) \}_{\tau=0}^{t-1}\right)) \nonumber \right.  \\
& \left.  \qquad \qquad \delta(\matr \gamma(t)-\matr h-\matr J \matr m(t))e^{i \matr \gamma (t)^\dagger\matr l(t)}\right\}.
\end{align}
Notice that $Z(\{\matr l(t)=\matr 0\})=1$. The statistics of the variables can be computed from the averaged generating functional $\left<Z(\{\matr l(t)\})\right>_{\matr J}$. In the large $N$ limit we obtain that (see~\ref{devDFT})
\begin{align}
\left< Z(\{\matr l(t)\})\right>_{\matr J}\simeq \prod_{n=1}^N\int{\rm d}\mathcal N(\{\phi_n (t)\}; 0,\mathcal C_{\phi})\prod_{t=0}^{T-1}\left\{{\rm d}m_n(t){\rm d}\gamma_n(t)\;{\delta}( m_n(t)-f_t\left\{m_n(\tau), \gamma_n(\tau)\right\}_{\tau=0}^{t-1})
\right. \nonumber \\
\left.  \delta\left(\gamma_n(t)-h_n-\sum_{s<t}\mathcal{\hat G}(t,s) m_n(s)-\phi_n(t)\right)e^{i\gamma_n(t){l}_n(t)} 
\right\}  \label{dftresult}
\end{align}
with $\mathcal N(\cdot;\mu,\Sigma)$ denoting the multivariate normal distribution with mean $\mu$ and covariance $\Sigma$. This result shows that in the large $N$ limit single trajectories can be treated as independent
following the effective stochastic dynamical process given by
\begin{align}
\matr m(t) &= f_t\left(\{\matr \gamma(\tau),\matr m(\tau)\}_{\tau=0}^{t-1}\right)\\
\matr \gamma(t)&=\matr h+\sum_{\tau=0}^{t-1}\mathcal{\hat G}(t,s) \matr m(\tau)+ \matr{\phi}(t) \ . \label{DFT1}
\end{align}
Here $\matr{\phi}(t)$ is a vector of independent Gaussian random variables with covariance matrix $\mathcal C_{\phi}$ given by 
\begin{equation}
\mathcal C_{\phi}= \sum_{n=1}^{\infty}c_n \sum_{k=0}^{n-2} \mathcal {G}^k\mathcal {C} (\mathcal {G}^\dagger)^{n-2-k} \ , \label{cov}
\end{equation}
where  $\mathcal{G}$ and $\mathcal{C}$ are $T\times T$ the response and the correlation matrices, respectively. With slight abuse of notation, their $(t+1,\tau+1)$ indexed entries are given by
\begin{align}
\mathcal G(t,\tau)&=\frac{1}{N} \sum_{i=1}^N\left<\frac{\partial m_i(t)}{\partial \phi_i(\tau)}\right>_{{\phi}_i} \label{response}\\
\mathcal C(t,\tau)&=\frac{1}{N} \sum_{i=1}^N\left<m_i(t)m_i(\tau)\right>_{\phi_i}.
\end{align}
Moreover the specific random matrix ensemble enters the result through the coefficients $c_n$, see \eqref{G3}, and  
the memory matrix $\mathcal {\hat G}$ given by 
\begin{equation}
\mathcal{\hat G}= {\rm R}(\mathcal{G}) \ .\label{mem}
\end{equation}
So far we have not yet referred to the TAP equations in the DFT analysis. Instead we have considered
a somewhat general  dynamical system with disorder and memory. Such a formulation gives us 
enough freedom to construct a convenient dynamics which asymptotically converges to the solution of the TAP equations.
We will define the dynamics to be of the form
\begin{equation}
\matr m(t+1)= \tanh(\matr \psi(t))
\label{update_Gauss}
\end{equation}
where the variables $\psi_i(t)$ must be chosen to become 
independent Gaussian fields in the resulting effective single variable
dynamics (\ref{DFT1}). In fact, there are actually various methods for doing so.
In the sequel we will limit our attention to a method that we call the {\em single step memory} construction.

\section{The Single Step Memory Construction}
In the single step memory algorithm we will construct the update 
 in such a way that the resulting memory term 
(\ref{mem}) satisfies the equation
\begin{equation}
\mathcal{\hat G}({t,\tau})= 0, \forall \tau\neq t-1 . \label{ssm}
\end{equation}
Hence, if \eqref{ssm} holds, then 
using \eqref{DFT1} we find that the variable
\begin{equation}
\matr \gamma(t)-\mathcal{\hat G}(t,t-1)\matr m(t-1) = \matr{\phi}(t) + \matr h
\end{equation}
becomes a a Gaussian field.
We will choose the field $\matr \psi (t)$ in \eqref{update_Gauss} as a linear combination of the Gaussian fields $\matr{\phi}(\tau)$, $\tau=1,\ldots,t$  of the form
\begin{equation}
\matr\psi(t)=\sum_{\tau=0}^{t}\mathcal A(t+1,\tau)(\matr{\phi}(\tau) + \matr h) \label{psi}
\end{equation}
where we have to construct the non-random terms ${\mathcal A}(t+1,\tau)$ to make the dynamical order parameters consistent with the single step memory condition \eqref{ssm}.
This condition leads to a very simple result for the response function \eqref{response} because there is no complicated
propagation in time of a response to an external field. In fact, from \eqref{update_Gauss} we obtain for the response function \eqref{response} 
\begin{align}
\mathcal G(t,\tau)&=\frac{1}{N}\sum_{i}\left\langle (1-m_i^2(t))\frac{\partial \psi_i(t-1)}{\partial \phi_i(\tau)}\right\rangle \\
&= (1-q(t))\mathcal A(t,\tau)
\end{align}
with $q(t)\triangleq\frac{1}{N} \matr m(t)^\dagger \matr m(t)$. Thus we have the explicit result
\begin{equation}
\mathcal A(t,\tau) =\frac{\mathcal G({t,\tau})}{1-q(t)}. \label{CoeffA}
\end{equation}
Finally, using \eqref{mem} we get an explicit result for the response function in terms of the memory terms $\mathcal{\hat G}(t,t-1)$. Note that by construction of the single step memory matrix $\mathcal {\hat G}$ \eqref{ssm}  
we can write \eqref{mem} as
\begin{equation}
\mathcal G({t,\tau})= a_{t-\tau}\prod_{s=\tau+1}^{t}\mathcal{\hat G}(s,s-1) \ ,
\end{equation}
where the coefficients $a_{n}$ are obtained from the power series expansion of the composition inverse of the R-transform:
\begin{equation}
{\rm R}^{-1}(x)=\sum_{n=1}^{\infty}a_{n}x^{n} \ . \label{inv}
\end{equation} 
By definition the trace of $\matr J$ is zero, i.e.\ $\rm R(0)=0$. Hence, the power series expansion in \eqref{inv} starts from the first order term. 

To complete the specification of the single step memory construction 
we only need to specify $\mathcal{\hat G}(t,t-1)$. This will be chosen  such that the method 
is asymptotically consistent with the static TAP equations. Specifically, from \eqref{Tap2} we should have 
\begin{equation}
\lim_{t\to \infty} \mathcal{\hat G}(t,t-1)= {\rm R}(1-q)\label{asym}{.}
\end{equation}
We choose the explicit form
\begin{equation}
\mathcal{\hat G}(t,t-1)=\frac{1-q(t)}{1-q(t-1)}{\rm R}(1-q(t-1)) \label{method1}.
\end{equation}
which assuming convergence $q(t)\to q$ as $t\to \infty$ leads to \eqref{asym}.
This form has also the advantage, that in (\ref{CoeffA}), for the update an unwanted factor $1- q(t+1)$, which would make $\matr\psi(t)$ depending on the future state $\matr m(t+1)$, cancels.

\subsection{Summary} 
Putting everything together the single-step memory algorithm for $t\geq 0$ is defined as
\begin{align}
\matr m(t+1)&= \tanh(\matr\psi(t)) \label{s1}\\
\matr \psi(t)&= Q(t)\sum_{\tau=0}^{t}a_{t+1-\tau} \matr u(\tau)\label{s2} \\
\matr u(t)&= \frac{\matr h+ \matr J \matr m(t)- \hat{\mathcal G}(t,t-1)\matr m(t-1)}{Q(t-1)(1-q(t))}\label{s3}
\end{align}
where we introduce
\begin{align}
Q(t)=\prod_{\tau=0}^{t}{\rm R}(1-q(\tau))=Q(t-1){\rm R}(1-q(t))
\end{align}
such that $Q(-1)=1$. The memory term $\hat{\mathcal G}(t,t-1)$ is given by \eqref{method1}. Moreover the algorithm initializes with $\matr m(t)=\matr 0$ for $t\in\{-1,0\}$.

\subsection{Asymptotic Consistency with TAP Equations}\label{conver}
In the sequel we show that if the single step memory algorithm \eqref{s1}-\eqref{s3} converges, it solves the TAP equations \eqref{Tap1}-\eqref{Tap2}. Let us assume that $\matr m(t)\to \matr m$ as $t$ tends to infinity. To have the convergence to the TAP equations we solely need to show that the sum in \eqref{s2} converges to
the proper limit. From \eqref{psi} and \eqref{CoeffA} we must have 
\begin{equation}
Q(t)\sum_{\tau=0}^{t}\frac{a_{t+1-\tau}}{(1-q(\tau))Q(\tau-1)} =\frac{1}{1-q(t+1)}\sum_{\tau=0}^{t} \mathcal G(t+1,\tau) \to 1.
\end{equation}
We make the so-called weak long-term response assumption \cite{Kurchan} that 
\begin{equation}
\lim_{t\to \infty}\mathcal G(t,\tau)=0, \quad \forall\text{ finite } \tau \label{wlt}.
\end{equation}
Hence, for sufficiently large $t$ and $\tau'<t$ such that $t/\tau'$ being finite as $t\to \infty$, we can write
\begin{align}
\sum_{\tau=0}^{t}\mathcal G(t+1,\tau) & \simeq\sum_{\tau=\tau'}^{t}\mathcal{G}(t+1,\tau )\label{con1}\\
& \simeq \sum_{\tau=\tau'}^{t}a_{t+1-\tau}{\rm R}\left(1-q\right)^{t+1-\tau}\\
&\simeq\sum_{n=1}^{\infty} a_{n}{\rm R}(1-q)^{n}\\
&= {\rm R}^{-1}({\rm R}(1-q))=1-q. 
\end{align}
Next we will provide the details of the single-step memory algorithm for the SK, Hopfield and random orthogonal models.

\subsection{Example~1 The SK-Model}
Recall that, for the standard SK model we have ${\rm R}(x)=\beta^2 x$; so that ${\rm R}^{-1}(x)=x/\beta^2$. Hence, $a_1=1/\beta^2$ and $a_n=0$ for $n>1$. Thus, the single-step memory algorithm may written as
\begin{align}
\matr m(t+1)&=\tanh(\matr \psi (t)) \label{Bolt1}\\
\matr \psi(t)&= \matr h+\matr J \matr m(t)-\beta^2(1-q(t))\matr m(t-1) \label{Bolt2}.
\end{align}
At first glance, these dynamical equations are similar but not exactly equal to those proposed by Bolthausen \cite{Bolthausen}. The difference is that instead of the dynamical order parameter $q(t)$ the fixed point solution of $q$ appears. 
Using the explicit form of the covariance of the field $\psi_i(t)$ given by \eqref{ssmcov} in the next section,
one finds for the field variance $\langle\left(\psi_i(t) - \langle \psi_i(t)\rangle\right)^2\rangle = \beta^2 q(t)$. Hence, if we start the iteration
(as in \cite{Bolthausen}) with $m_i(1) = \sqrt{q}$ such that  $q(1) = q$, then we find that in the large $N$ limit, we also have $q(t) = \left\langle \tanh^2(\psi_i(t-1)\right\rangle = q$ for all times $t$ and we get agreement with \cite{Bolthausen}.

\subsection{Example~2 The Hopfield Model}
For the Hopfield model from \eqref{hopfield} we have
\begin{equation}
{\rm R}^{-1}(x)=\frac{1}{\beta \alpha}\frac{x}{\beta-x}.
\end{equation}
Thus the memory coefficients are given as $a_n=1/(\alpha \beta^{n+1})$ for $n\geq 1$. In the sequel we show that the single step memory algorithm for the Hopfield model coincides with AMP algorithm which was introduced in the context of the CDMA 
problem in \cite{Kabashima} and compressed sensing in \cite{Donoha}. From \eqref{s2} we first write
\begin{align}
\matr \psi(t)&= \frac{Q(t)}{\alpha \beta^2}\matr u(t)+ \frac{Q(t)}{\alpha \beta^{t+2}} \sum_{\tau=0}^{t-1} \beta^\tau \matr u(\tau)\\
&=\frac{Q(t)}{\alpha \beta^2}\matr u(t)+  \frac{1}{\beta}{\rm R}(1-q(t))\matr \psi(t-1). \label{keyamp}
\end{align}
For convenience let us introduce 
\begin{equation}
A(t)\triangleq \frac{{\rm R}(1-q(t))}{\beta\alpha(1-q(t))}=\frac{\beta}{1+\beta\alpha(1-q(t))}.
\end{equation} 
Notice that from \eqref{s3} we may write \eqref{keyamp} in the form of
\begin{align}
\matr \psi (t)=\frac{1}{\beta}A(t)[\matr h+\matr J \matr m(t)]+\alpha(1-q(t))A(t)[\matr \psi(t-1)- A(t-1)\matr m(t-1)].
\end{align}
Then, defining $\matr z(t) \triangleq \matr \psi(t)- A(t)\matr m(t)$, we write the single step memory algorithm as
\begin{align}
\matr m(t+1)&=\tanh(\matr z (t)+A(t) \matr m(t))\\
\matr z(t)&=\frac{1}{\beta}A(t)[\matr h+(\matr {J}-\beta\matr {\bf I}) \matr m(t)]+\alpha(1-q(t))A(t)\matr z(t-1)
\end{align}
where $\matr {\bf I}$ is the identity matrix of appropriate dimension. Note, that $(\matr {\bf I}-\matr J/\beta)$ asymptotically coincides with the corresponding central Wishart matrix (see Section~\ref{GIRME}). Thereby we exactly obtain the AMP iteration steps as introduced in \cite{Kabashima}. We also refer the reader to the related works \cite{Bayati} and \cite{Mimura}, where the dynamics of the AMP algorithm is analyzed by means of DFT and Bolthausen's conditioning technique \cite{Bolthausen}, respectively. 

Bolthausen's conditioning technique for SK model \cite{Bolthausen} and Hopfield model \cite{Bayati} are based on the assumption that the entries of the underlying coupling matrix are defined via zero-mean iid and \emph{Gaussian distributed} random variables, see {Section~\ref{Rtrasform}}. Recently, it has been shown in \cite{bayati15} that the same analyses can be obtained without the need of \emph{Gaussian distribution} assumption but a sub-Gaussian tail condition of the distribution is required. Indeed, thanks to the central limit theorem, one can show that the generating function (\ref{G1}) also yields the same result regardless of whether the \emph{Gaussian distribution} assumption is considered or not, see \cite[Section~5]{Tanaka08}.

\subsection{Example~3 The Random Orthogonal Model}
For the random orthogonal model from \eqref{ROM} we have ${\rm R}^{-1}(x)=x/(\beta^2-x^2)$. This yields the memory coefficients as 
\begin{equation}
a_n= \left\{\begin{array}{ll}
\frac{1}{\beta^{n+1}}&  \text {$n$ is odd}\\
0                & \text {$n$ is even}.
\end{array}\right.
\end{equation}

In Figure~1 and Figure~2
\begin{figure}
\centering 
\includegraphics[height=3.8in,width=5in]{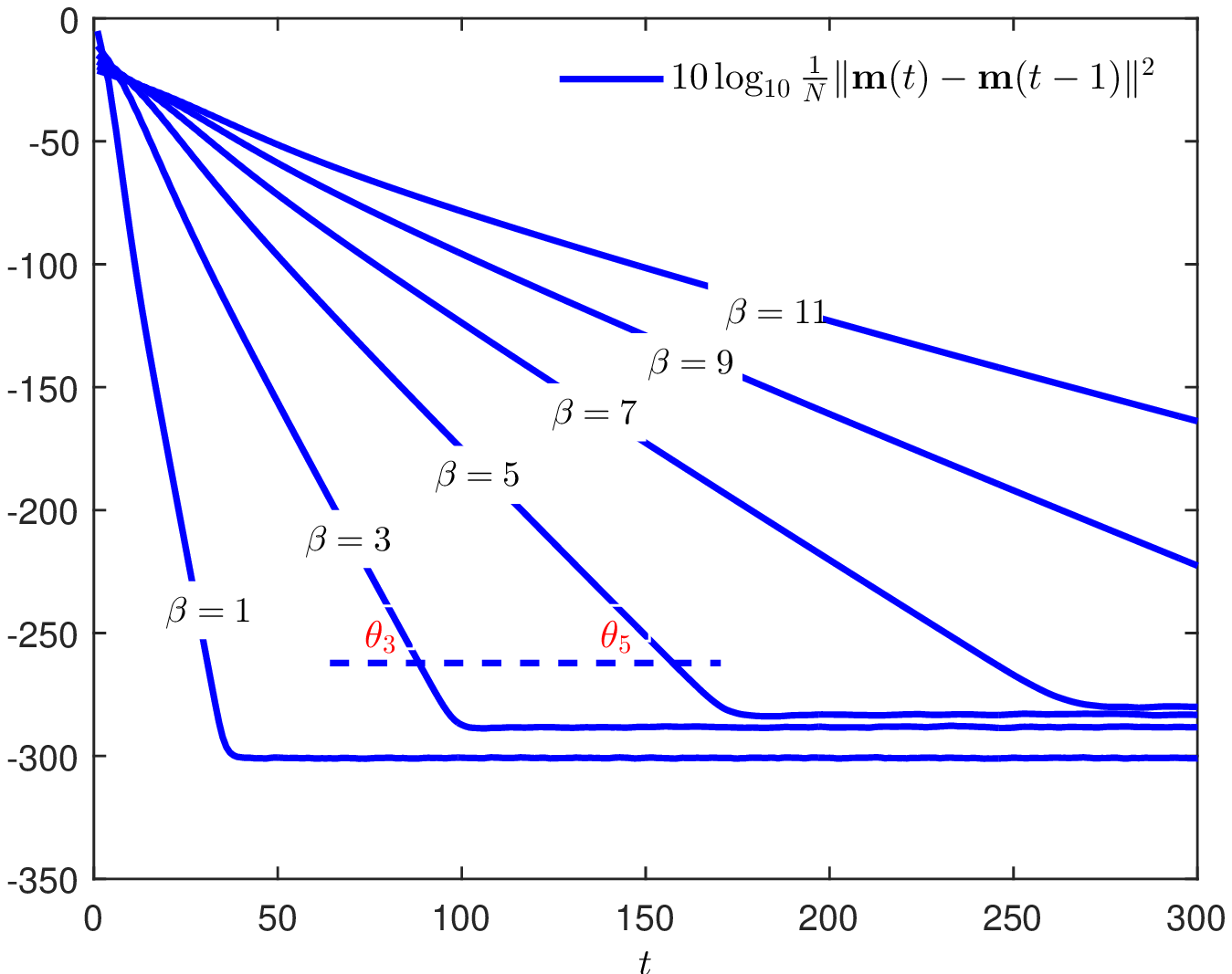}
\caption{Random orthogonal model with $\beta\in \{1,3,...,9,11\}$, $h_i= 1$ and $N=2^{14}$. Here e.g. $\theta \in \{\theta_3, \theta_5\}$ substitutes the respective linear decay time (in the log-domain), i.e. the convergence is faster bigger $\theta$ is.}
\vspace{1cm}
\includegraphics[height=3.8in,width=5in]{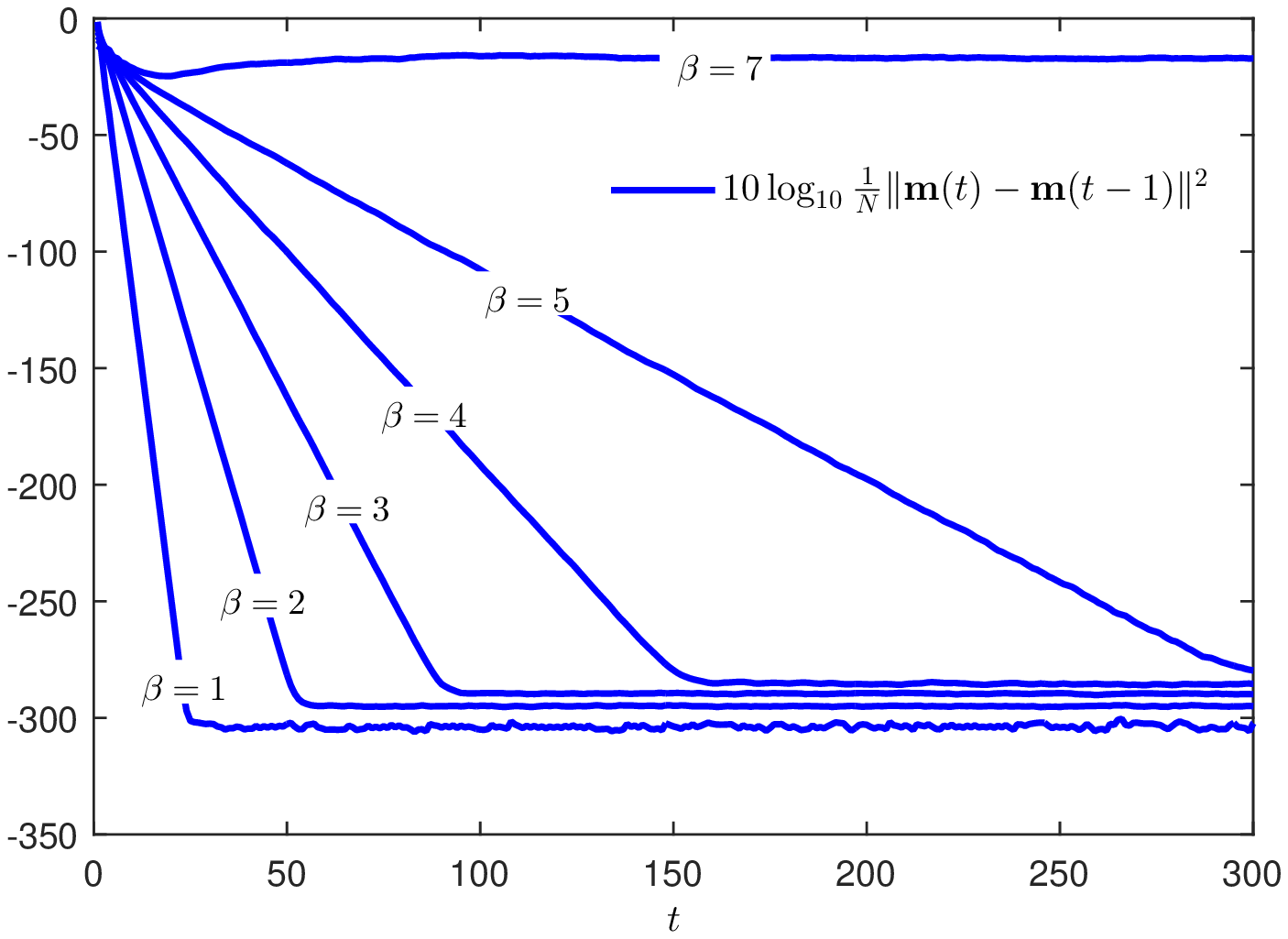}
\caption{Random orthogonal model with $\beta\in \{1,...,5,7\}$, $h_i= 2$ and $N=2^{14}$. The inverse temperature $\beta=6.9$ gives the AT line.}
\end{figure}
we illustrate the convergence of the single-step memory algorithm for the random orthogonal model
obtained by running simulations. Notice that after few iteration steps the convergence becomes exponentially fast.  The flat lines around $(-300)$dB, i.e. $10^{-30}$, are the consequence of the machine precision of the computer which was used. 

The convergence improves with increasing temperature parameter $1/\beta$. 
On the other hand, for large enough $\beta$, the algorithm fails to converge. To estimate the critical parameter, we study
the inverse decay time measured by the angle $\theta$ as illustrated in Figure~1 and extrapolate the simulational data
to $\theta=0$ using a convenient range of $\beta$. One might expect that the critical $\beta$ would coincide with the one obtained from a de Almeida--Thouless (AT) stability condition which can also be derived from the TAP approach \cite{adatap}.
The AT line is given by the equation
\begin{equation}
\alpha{\rm R'}(1-q)=1 \quad \text{ with} \quad  \alpha\triangleq\frac{1}{N}\sum_{i=1}^{N} \left<(1-\tanh^2(\psi_i))^2\right> \label{AT}
\end{equation}
where the random variable $\psi_i$ is a Gaussian with mean $h_i$ and variance $q{\rm R}'(1-q)$. Note, that \cite{adatap} contains a typo in the corresponding expression.
In Figure~3, 
\begin{figure}
\centering 
\includegraphics[height=4in,width=5in]{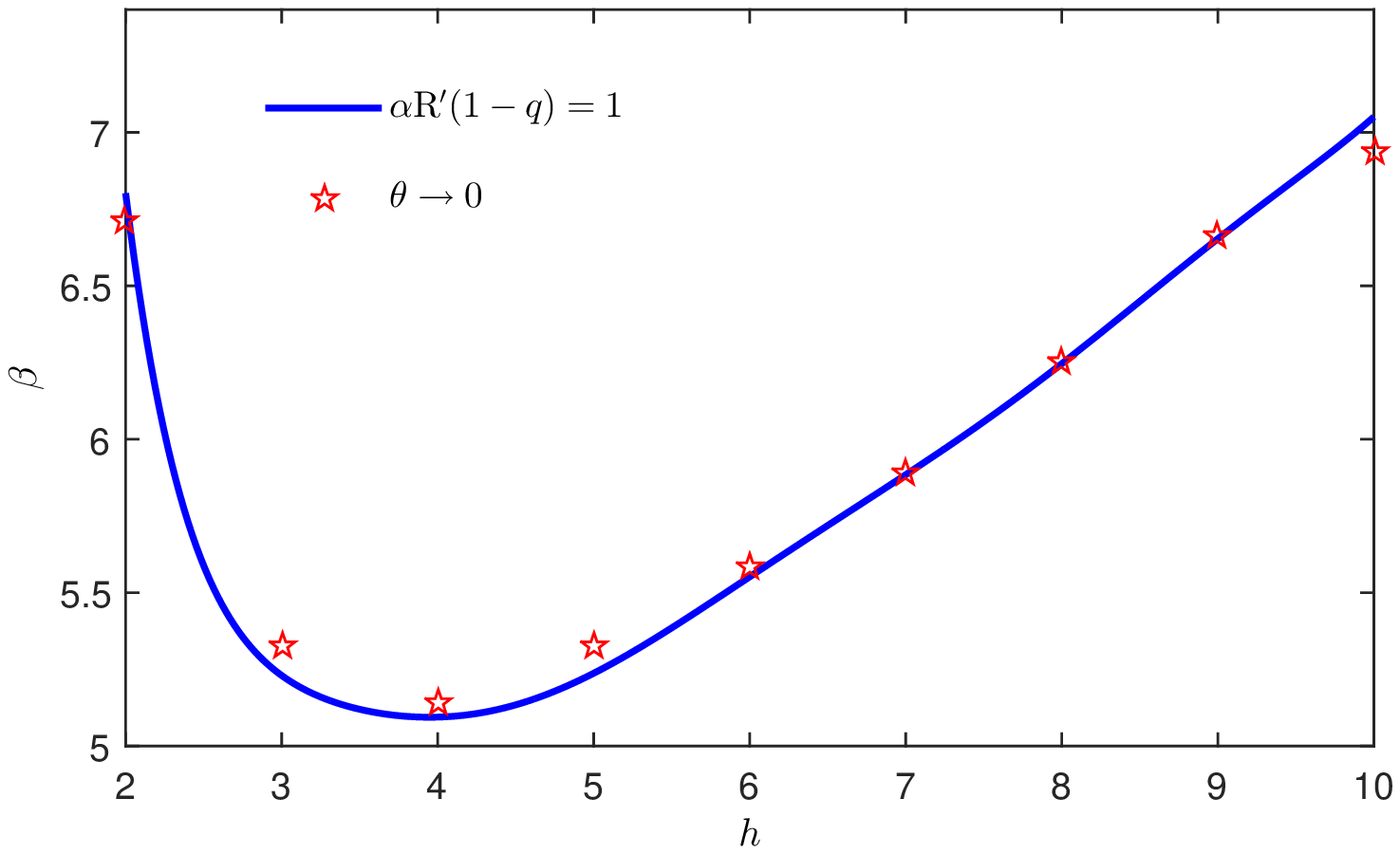}
\caption{Consistency of the diverging decay time with the AT line: The simulations were based on the average over ten realizations of $\matr J$ with $N=2^{14}$.}
\end{figure}
we present a comparison between the simulations and \eqref{AT}. This coincidence can be understood from a dynamical point of view by analyzing the stability of the dynamics close to the fixed point. The details will be postponed to Section~V. Finally, it also worth noting that the trajectories of the algorithm show a self-averaging behaviour above the AT line for large $N$. On the other hand, we find that below the AT line, there are strong sample to sample fluctuations. However, by averaging order parameters over many samples, we get a good agreement with the theory. Since the main goal of this paper is to present a convergent algorithm we will leave a more careful investigation of this point to future publications.

\subsection{The Field Covariance Matrix}
In order to compare simulations of systems with analytical results obtained from 
the dynamical functional approach in the limit $N\to\infty$ and to study the stability of TAP
fixed--points we have to perform expectations over the
Gaussian random variables $\psi_i (t)$ (see \eqref{psi}). Specifically we write
\begin{equation}
	\frac{1}{N}\langle\Vert \matr m(t)-\matr m(t-1)\Vert^2\rangle_{\matr J} =  q(t)+q(t-1)-2\mathcal C(t,t-1)
\end{equation}
where the order parameter $\mathcal C$ is given by
\begin{equation}
	\mathcal C(t+1,t'+1)= \frac{1}{N}\sum_{i=1}^{N}\int {\rm dP}(x,y)\; \tanh (\langle\psi_i(t)\rangle+x)\tanh (\langle \psi_i(t')\rangle+y).
\end{equation}
Here ${\rm P}$ denotes a two-dimensional Gaussian distribution with zero mean.  The mean of the field $\psi_i(t)$ follows from \eqref{psi} as
\begin{equation}
	\langle\psi_i(t) \rangle= \left\{Q(t) \sum_{\tau=0}^{t}\frac{a_{t+1-\tau}}{(1-q(\tau))Q(\tau-1)}\right\}h_i.
\end{equation} 
Hence, we need to compute the corresponding covariance matrix which is defined as
\begin{equation}
	\mathcal C_{\psi}(t,t')= \left<(\psi_i (t)-\left<\psi_i(t)\right>)(\psi_i (t')-\left<\psi_i(t')\right>)\right>. \label{covf}
\end{equation}
In \ref{proofssmcov} we derive the expression 
\begin{equation}
	\mathcal C_{\psi}(t,t')= Q(t)Q(t')\sum_{l\leq t, m\leq t'}\frac{{\rm  Co}_{x^{t+1-l}y^{t'+1-m}}[{\rm A}(x,y)]\mathcal C(l,m)}{(1-q(l))(1-q(m))Q(l-1)Q(m-1)}.  \label{ssmcov}
\end{equation}
In this expression, for a power series $f(x,y)=\sum_{n,k\geq 0}a_nb_k x^ny^k$, we have introduced the symbol ${\rm Co}_{x^ny^k} [f(x,y)]\triangleq a_n b_k$ for its coefficients. Moreover the function $\rm A$ is defined as
\begin{equation}
	{\rm A}(x,y)\triangleq \left(\frac{1}{{\rm R}^{-1}(x)}-\frac{1}{{\rm R}^{-1}(y)}\right)^{-1}(y-x) \label{Afunction}.
\end{equation}

The function ${\rm A}$ has a relatively simple form for the three random matrix ensembles considered in this paper. For the SK and Hopfield models we have ${\rm A}(x,y)=xy/\beta^2$ and ${\rm A}(x,y)=xy/(\beta^2\alpha)$, respectively. Moreover, for the random orthogonal model, from ${\rm R}^{-1}(x)=x/(\beta^2-x^2)$ we have ${\rm Co}_{x^ny^k} [{\rm A}(x,y)]=\delta_{nk}(-1)^{n+1}/\beta^{2n}$. 
 We next compare our simulations with theoretical results. We used the initializations $\matr m(t)=\matr 0$ for $t\in \{0,1\}$, hence we assign $\mathcal C(1,0)=0$. In Figure~4 and 5 
\begin{figure}
	\centering 
	\includegraphics[height=4in,width=5in]{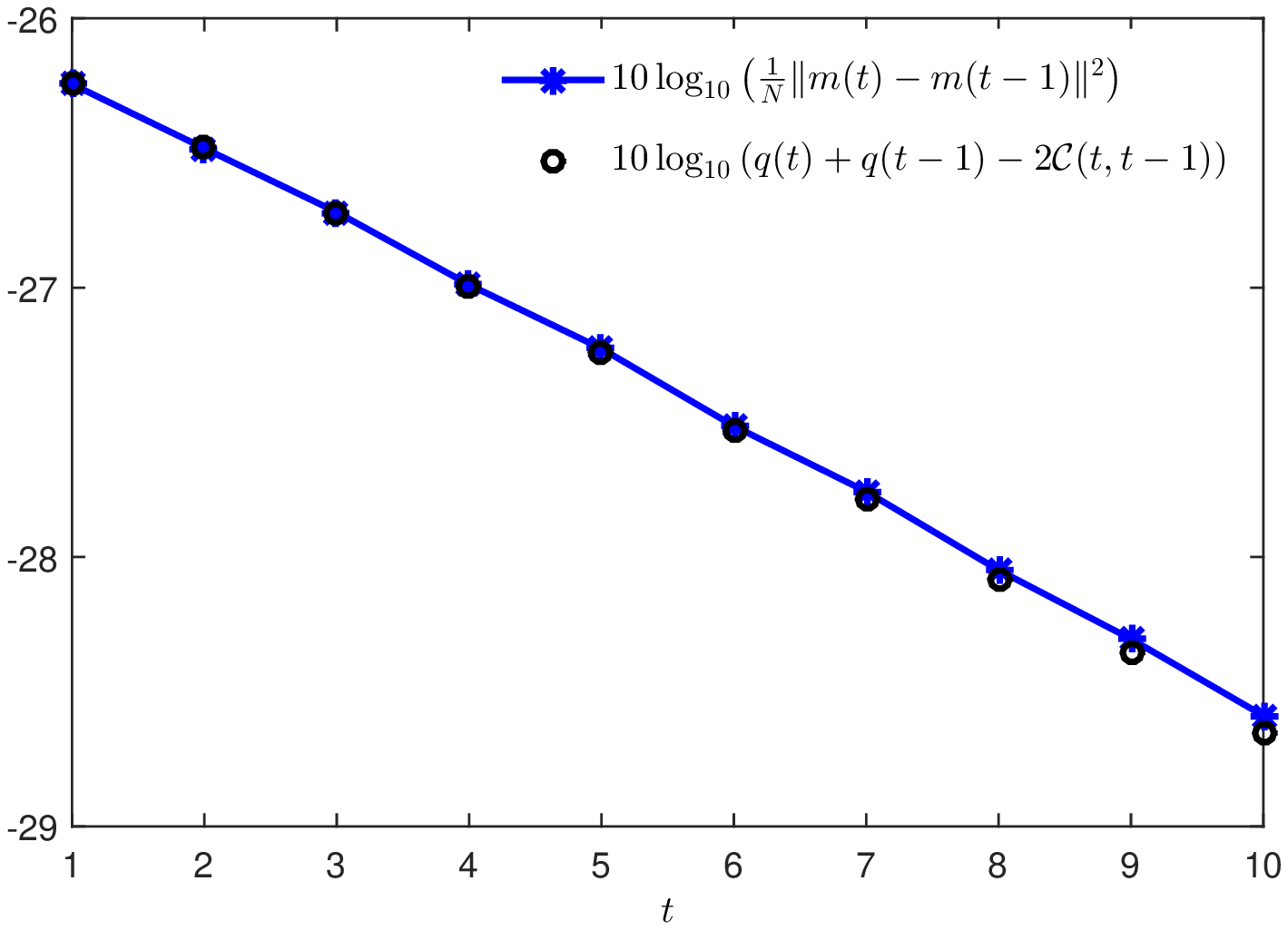}
	\caption{Above AT line: Comparison of the theory and simulation for $\beta=20$ and $h_i= 1$, $N=2^{14}$.} 
	\includegraphics[height=4in,width=5in]{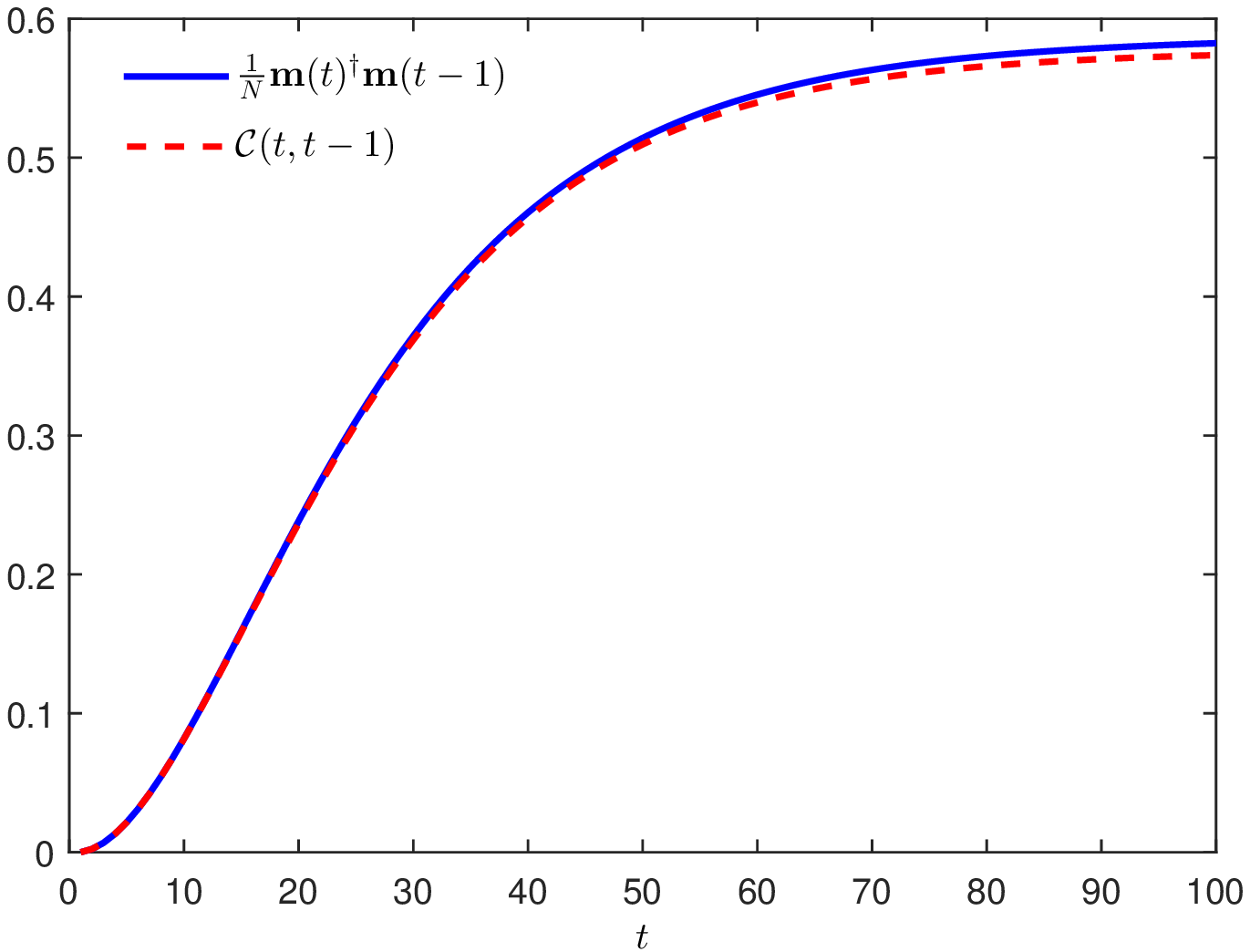}
	\caption{Above AT line: Comparison of the theory and simulation for  $\beta=20$ and $h_i= 1$, $N=2^{14}$.} 
\end{figure}
we show such a comparison above the AT line. Note, that no averaging over coupling matrices was used for the simulations. The  integration over two-dimensional correlated Gaussian distribution used to calculate $\mathcal C(t,t-1)$ was performed  numerically. However the accuracy of the numerical method limits us for providing very precise results as $t$ grows. In 
Figure~4 we illustrate the theoretical prediction of $\frac{1}{N}\matr m(t)^\dagger \matr m(m-1)$ by the order parameter $\mathcal C(t,t-1)$ for a large range of $t$.  Below the AT line,
\begin{figure}
	\centering
	\includegraphics[height=4in,width=5in]{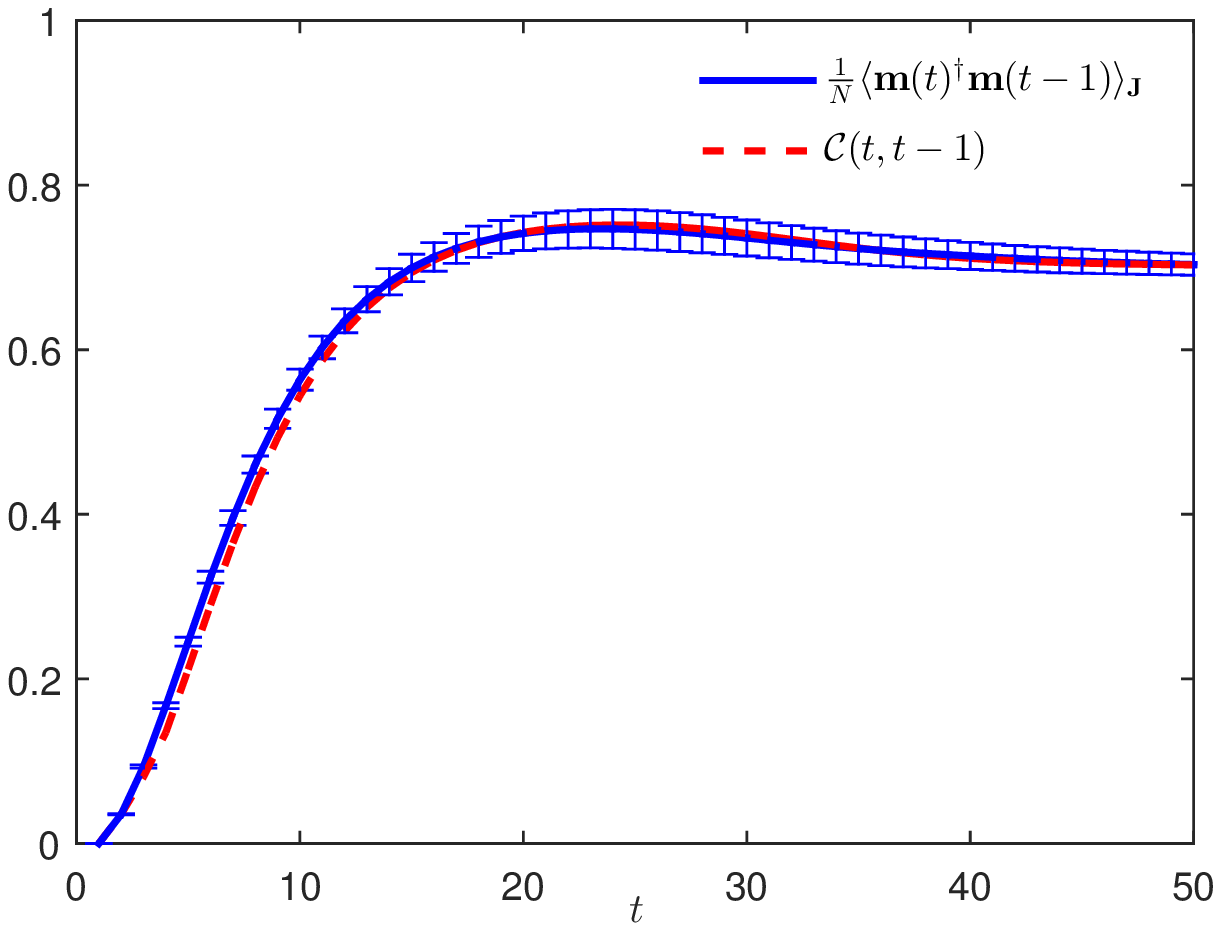}
	\caption{Below AT line: Comparison of the theory and simulation for $\beta=10$ and $h_i= 2$, $N=2^{12}$. $\left< \cdot \right>_{\matr J}$ is obtained by $5\times 10^3$ realizations of $\matr J$.} 
\end{figure}
the single-step memory algorithm diverges and simulated trajectories show strong sample fluctuations. However, by taking an average over a large number 
of trajectories we obtain a good agreement with the theory (see Figure~6).

\subsection{Asymptotic consistency with Cavity Variance}
In Section~\ref{conver} we have demonstrated the convergence of the single step memory algorithm to the TAP equations. In a similar way one can show that \eqref{ssmcov} converges to the variance of the static field variance in \eqref{rs} as $t$ and $t'$ tend to infinity. Specifically,  by invoking the weak long-term response assumption \eqref{wlt} in \eqref{cov} and following similar steps 
as in  \eqref{proofssmcov}, for sufficiently large $t$, $\tau<t$, $t'$ and $\tau'<t'$ such that $\tau/t$ and $\tau'/t'$ being finite as $t$ and $t'$ tend infinity, one can show that
\begin{align}
\mathcal C_{\psi}(t,t')&\simeq \frac{q}{(1-q)^2}\sum_{\tau\leq t, \tau'\leq t'}{\rm  Co}_{x^{t+1-\tau}y^{t'+1-\tau'}}[{\rm A}(x,y)]{\rm R}(1-q)^{t+1-\tau} {\rm R}(1-q)^{t'+1-\tau'}\label{keybegin}\\
&\simeq  \frac{q}{(1-q)^2}\sum_{n,k\leq 1}{\rm  Co}_{x^{n}y^{k}}[{\rm A}(x,y)]{\rm R}(1-q)^{n}{\rm R}(1-q)^{k} \\
& =\frac{q}{(1-q)^2}{\rm A}({\rm R}(1-q),{\rm R}(1-q)). \label{keyend} 
\end{align}
In this expression, by abuse of notation, we denote $\lim_{y\to x}{\rm A}(x,y)$ by ${\rm A}(x,x)$. Its explicit form is given by
\begin{equation}
\lim_{y\to x}{\rm A}(x,y)=({\rm R}^{-1}(x))^2\frac{1}{({\rm R}^{-1}(x))'}=({\rm R}^{-1}(x))^2{\rm R}^{'}({\rm R}^{-1}(x)). \label{limitA}
\end{equation}
Hence we have 
\begin{align}
\lim_{t,t'\to \infty}\mathcal C_{\psi}(t,t')&= \frac{q}{(1-q)^2}({\rm R}^{-1}({\rm R}(1-q)))^2{\rm R}^{'}({\rm R}^{-1}({\rm R}(1-q))) \\
&= q{\rm R}'(1-q).
\end{align}

\section{The stability of TAP fixed points}
In order to analyze the stability of the fixed points of the single step algorithm  we resort to a linear stability analysis. 
We add Gaussian white noise to the dynamics, i.e.\ we set $\psi_i(t) \rightarrow \psi_i(t) + \epsilon_i(t)$ with $\langle\epsilon_i(t)^2\rangle = \epsilon$ and discuss the limit $\epsilon\to 0$. If the static TAP fixed point is stable, then the system should asymptotically show
only small stationary fluctuations around this and we can work in the Fourier domain. Hence, we assume 
\begin{align}
\mathcal C(t,t')&=\frac{1}{2\pi} \int {\rm d}\omega\; \hat{\mathcal C}(\omega) e^{i \omega (t-t')}\\
\mathcal C_{\psi}(t,t')&= \frac{1}{2\pi}\int {\rm d}\omega\;\hat{\mathcal C}_{\psi}(\omega) e^{i \omega(t-t')}.
\end{align}
Inserting these Fourier representations into \eqref{ssmcov}, for large $t$ and $t'$ we may write (see \eqref{keybegin}-\eqref{keyend})
\begin{align}
 \hat{\mathcal C}_{\psi}(\omega)& \simeq \sum_{l\leq t,m\leq t'}\frac{{\rm Co}_{x^{t'+1-l}y^{t'+1-m}}[{\rm A}(x,y)]\hat{\mathcal C}(\omega) 
 e^{i \omega (l-t -(m-t'))}}{(1-q)^2 {\rm R}(1-q)^{l-t-1}{\rm R}(1-q)^{m-t'-1}}\\
& \simeq \frac{{\rm A}(e^{-i \omega}{\rm R}(1-q),e^{i\omega}{\rm R}(1-q) )}{(1-q)^2}\mathcal{\hat{C}}(\omega). \label{covkey}
\end{align}
For small noise $\epsilon \to 0$, the assumption of stability translates into small fluctuations around the static solution 
and we can write
\begin{align}
\hat{\mathcal C}(\omega)&\simeq  2\pi q\delta(\omega)+\epsilon\hat{c}(\omega)\\
\hat{\mathcal C}_{\psi}(\omega)&\simeq 2\pi q{\rm R}'(1-q)\delta(\omega)+\epsilon\hat{c}_{\psi}(\omega).
\end{align}
where we have separated fluctuations into static and dynamical parts. We will analyse the dynamical part 
next, but note, that also the static 
part $q$ will have contributions from $\epsilon$. Thus for $\omega\neq 0$ we have
\begin{equation}
\hat{c}_{\psi}(\omega)=\hat c(\omega)\frac{{\rm A}(e^{-i \omega}{\rm R}(1-q),e^{i\omega}{\rm R}(1-q) )}{(1-q)^2}.\label{AT1}
\end{equation}
where now the value of $q$ is computed for $\epsilon = 0$. We next express $\hat{c}(\omega)$ in terms of $\hat{c}_{\psi}(\omega)$ 
for small $\epsilon$. The calculation in \ref{ATdetails} is based on expanding
\begin{equation}
\mathcal C(t+1,t'+1)=\frac{1}{N} \sum_i \left\langle \tanh(u(t) + h_i) \tanh(u(t') + h_i) \right\rangle_{u}
\end{equation}
up to first order in $\epsilon$. The brackets denote expectations over the two dimensional Gaussian field 
$(u(t), u(t'))$ with $\langle u(t) u(t')\rangle \simeq s_0 + \epsilon (s(t - t') + \delta_{t,t'})$ for $\epsilon\to 0$, where
$s_0 = q{\rm R}'(1-q)$ and $s(t - t') = c_\psi(t,t')$.
For $t=t'$ the integral is over a single Gaussian only.
The calculation shows that
\begin{equation}
\hat{c}(\omega) = \alpha(1+\hat{c}_{\psi}(\omega))
\label{comega}
\end{equation}
with $\alpha$ is defined as in \eqref{AT}. Combining this relationship with \eqref{AT1} we have
\begin{equation}
\hat{c}(\omega)\simeq\alpha\left(1-\frac{\alpha {\rm A}(e^{-i \omega}{\rm R}(1-q),e^{i\omega}{\rm R}(1-q) )}{(1-q)^2} \right)^{-1}.
\label{dynfluctuat}
\end{equation}
In fact, for the SK, Hopfield and random orthogonal models, we have ${\rm Co}_{x^ny^k} [{\rm A}(x,y)]=0$ $\forall n\neq k$. Therefore \eqref{dynfluctuat} is actually independent of $\omega$ and from \eqref{limitA} we explicitly have that
\begin{equation}
\hat{c}(\omega)=\frac{\alpha}{1-\alpha{\rm R}'(1-q)}.
\end{equation}

In general, the right hand side  of \eqref{dynfluctuat} must be non--negative to have a valid representation as a Fourier-transform of a time dependent
correlation function. While the term ${\rm A}(e^{-i \omega}{\rm R}(1-q),e^{i\omega}{\rm R}(1-q) )$ is
always positive, see \eqref{covkey}, the second term is small and positive for sufficiently small $\beta$. But it
changes sign and diverges. One expects that the divergence will occur first for the long range fluctuations, i.e.\ for
the limit of low frequencies.  Taking the limit yields
\begin{equation}
\lim_{\omega\to 0}\hat{c}(\omega)=\frac{\alpha}{1-\alpha{\rm R}'(1-q)}.
\end{equation}
The condition
\begin{equation}
\alpha{\rm R}'(1-q)  = 1
\end{equation}
for the onset of instability agrees with the well--known AT stability criterion \cite{adatap}.

\section{Discussion and Outlook} 
In this paper we have presented a theoretical approach to the design of iterative algorithms 
for solving the TAP equations for Ising models with random couplings drawn from general invariant
ensembles. We were guided by the idea that one needs to subtract terms from the
internal field which depend on the values of the magnetizations at previous times. Using 
dynamical functional theory we have shown that in such a way, memory terms can be canceled and one arrives
at a Gaussian distributed field, which eventually converges to the cavity field provided that a
stability condition is fulfilled. We have presented a specific
method which we have called the 'Single Step Memory Construction'. Our approach may be extended
in several ways. For example other 
subtraction methods are possible. One might design an alternative scheme, where the response function is 
required to be zero after one time step leading to a somewhat different algorithm and we will give details 
elsewhere. It would be interesting to see in which cases the explicit memory terms in the subtraction
method can be simplified by introducing auxiliary variables as is possible for the Hopfield model.
Other extensions of our method would be to more general probabilistic models beyond the simple Ising case.
This would include continuous random variables and other forms of interactions. An application to models
of compressed sensing would be interesting where certain random matrix ensembles (such as the random orthogonal ones)
might be natural models for the observation matrix. Specifically one can trivially extend the random orthogonal ensemble by considering the more general spectrum such that the eigenvalues of $\matr {\tilde J}$ are distributed as $\alpha \delta(\lambda-1)+(1-\alpha) \delta(\lambda+1)$. In the context of compressed sensing this model coincides with the so-called random row-orthogonal ensemble
\cite{Mikko}, \cite{Tulino13}. We will discuss details 
in a forthcoming publication.  
Finally, it would be important to address a drawback of our method which prevents an application to 
probabilistic inference problems with arbitrary data. Our subtraction scheme depends explicitly on the random matrix 
ensemble of couplings which may not be known in practice. Hence it would be interesting to develop
schemes which adapt to the concrete data which would then achieve convergence to `adaptive TAP equations' 
of \cite{adatap}, providing possible alternatives to the currently applied message passing algorithms \cite{Minka1}. 

\section*{Acknowledgment}
The authors would like to thank Florent Krzakala for inspiring us to do this study. This work was partially supported by the European Commission in the framework of the FP7 Network of Excellence in Wireless COMmunications NEWCOM$\sharp$ (Grant agreement no. 318306). 

\appendix
\section{The proof of ${\tilde J}_{ii}-\frac{1}{N}{\tr}(\matr {\tilde J})\to 0$} \label{Profof2}
From \eqref{spectral} we have 
\begin{equation}
\tilde J_{ii}= \sum_{n=1}^{N}\lambda_n [(\matr O_{n}\matr O_{n}^\dagger)]_{ii}=\sum_{n=1}^{N}\lambda_nO_{in}^2
\end{equation}
where $\matr O_n$ the $n$th column vector of Haar (orthogonal) $\matr O$ and $\lambda_n=\Lambda_{nn}$. Note that $\matr \Lambda$ is independent of $\matr O$. For convenience we treat the realizations of $\matr\Lambda$ from its probability space and denote as $\matr\Lambda(\omega)$. We show the convergence for every realization $\tilde J_{ii}(\omega)=\sum_{n=1}^{N}\lambda_n(\omega)O_{in}^2$.  The mean and variance of $\tilde J_{ii}(\omega)$ are respectively given by
\begin{align}
\left< \tilde J_{ii}(\omega)\right>&=\sum_{n=1}^{N}\lambda_n(\omega) \left<O_{in}^2\right> \\
{\rm Var}[\tilde J_{ii}(\omega)]&=\sum_{n=1}^{N}\lambda^2_n(\omega){\rm Var}[O_{in}^2]+2\sum_{n<k}\lambda_n(\omega)\lambda_k(\omega) {\rm Cov}[O_{in}^2,O_{ik}^2] \label{Var}
\end{align}
where for the random variables $X$ and $Y$, ${\rm Var}[X]$ and ${\rm Cov}[X,Y]$ denoting the variance $X$ and the covariance of $X$ and $Y$, respectively. For the proof, we basically need to show that
\begin{equation}
\lim_{N\to \infty}{\rm Var}[\tilde J_{ii}(\omega)]=0. \label{suf.}
\end{equation}
To do this we make use of the so-called (orthogonal) Weingarten calculus that allows to for calculate
 joint moments of Haar entries (analogeous to Wick calculus for Gaussian matrices). For details we refer the reader to \cite{Collins-a, Collins-b}. From \cite[Theorem~2.1, {Example~2.1}]{Collins-b} {we have}
\begin{equation}
\left<O_{in}^2O_{ik}^2\right>=\left\{\begin{array}{cc}
\frac{{N+1}}{N(N+2){(N-1)}} & n\neq k\\
\frac{3}{N(N+2)} & n=k 
\end{array}\right. .
\end{equation}
Furthermore, we have $\left<O_{ij}^2\right>=1/N$, $\forall i,j$. Thus, the variance \eqref{Var} reads
\begin{equation}
{\rm Var}[\tilde J_{ii}(\omega)]=\frac{{2(N-1)}}{N^2(N+2)}\sum_{n=1}^{N}\lambda^2_n(\omega) { +} \frac{4}{N^2(N+2){(N-1)}}\sum_{n<k}\lambda_n(\omega)\lambda_k(\omega). \label{vafin}
\end{equation}
Note that the spectrum $\matr {\tilde J}$ is assumed to converge almost surely to a compactly supported probability distribution such that the smallest and largest eigenvalue of $\matr {\tilde J}$ converge (almost surely) to the infimum and supremum of the compact support, respectively. This implies that the minimum and maximum of the eigenvalues of $\matr {\tilde J}$ are uniformly bounded above for a sufficiently large $N$. This is a sufficient to get \eqref{suf.} from \eqref{vafin}. Thereby we complete the proof.

\section{The self-averaging limit of adaptive TAP Equations}\label{derTAP}
We will provide a derivation of the TAP equations \eqref{Tap1}-\eqref{Tap2} 
from the `adaptive TAP' approach of \cite{adatap}. Under the assumption of 
Gaussian distributed cavity fields and an approximate linear response argument one finds
\begin{align}
m_i&= \tanh\left(h_i+ \sum_{j} J_{ij}m_j-V_im_i\right)\\
V_i&= \Lambda_{ii}-\frac{1}{\chi_{ii}} \label{cav}\\
\Lambda_{ii} &=V_i+\frac{1}{1-m_i^2}
\end{align}
with the positive definite matrix $\matr \chi\triangleq (\matr \Lambda-\matr J)^{-1}$.  

To obtain the TAP equations in \eqref{Tap1}-\eqref{Tap2} we basically need to show that $V_i\simeq{\rm R}(1-q)$. To that end we write \eqref{cav} in the form of  
\begin{equation}
V_i= \Lambda_{ii}-\left(\frac{\partial \ln \det (\matr \Lambda-\matr J)}{\partial \Lambda_{ii}}\right)^{-1}. \label{trick} 
\end{equation}
Our basic idea is to simplify $\ln \det (\matr \Lambda-\matr J)$ using results of {\em free probability theory} for random matrices.  To that end we invoke an additional assumption that {the empirical distribution function of $\{\Lambda_{11},\cdots,\Lambda_{NN}\}\in\RR^N$ converges weakly and almost surely to a compactly supported probability distribution as $N\to \infty$}. Since $\matr J$ is asymptotically invariant and having a compactly supported limiting spectrum, $\matr \Lambda$ and $\matr J$ are asymptotically (almost surely) free \cite{Collins-a}. Thus, due to the uniform convergence property of the R-transform, see \cite[Lemma~3.3.4]{Hiai}, for a sufficiently large $N$ we have
\begin{align}
{\rm R}_{\matr \Lambda-\matr J}^N(x)&\simeq {\rm R}_{\matr \Lambda}^N(x)+{\rm R}_{-\matr J}^N(x)\label{addf}\\
&= {{\rm R}_{\matr \Lambda}^N(x)-{\rm R}_{\matr J}^N(-x)}
\end{align}
where we denote the R-transform of the spectrum of an $N\times N$ symmetric matrix $\matr X$ by ${\rm R}_{\matr X}^N$. Note that we have  $1-q=\frac{1}{N}{\rm tr}(\matr \Lambda-\matr J)^{-1}$. Then, from Lemma~1 below we have
\begin{align}
\frac{1}{N}\ln \det (\matr \Lambda-\matr J)&= -(1+\ln (1-q))+ \int_{0}^{1-q}{\rm d}\omega\;{\rm R}^N_{\matr \Lambda-\matr J}(-\omega) \label{dev1}\\
&\simeq  -(1+\ln (1-q))+\int_{0}^{1-q}{\rm d}\omega\; {\rm R}^N_{\matr \Lambda}(-\omega)-\int_{0}^{1-q} {\rm d}\omega\;{\rm R}^N_{\matr J}(\omega) \label{dev2} \\
&= \frac{1}{N}\ln \det (\matr \Lambda-V\matr {\bf I})+(1-q)V-\int_{0}^{1-q} {\rm d}\omega\; {\rm R}^N_{\matr J}(\omega) \label{dev3}
\end{align}
where $V$ is defined through the implicit equation $(1-q)=\frac{1}{N}\sum_{i=1}^N\frac{1}{\Lambda_{ii}-V}$. This implies that ${\rm tr}(\matr \Lambda-\matr J)^{-1}\simeq {\rm tr}(\matr \Lambda-V\matr {\bf I})^{-1}$. Here, from \eqref{dev2} to \eqref{dev3} we make use of the identity \eqref{nice2} below.  Note that for a non-negative $N\times N$ matrix $\matr X$ we can write the Stieltjes transform of the eigenvalue distribution of $\matr X$, say ${\rm P}^N_{\matr X}$, as ${\rm M}^N_{\matr X}(\omega)=\int {\rm dP}^N_{\matr X}(x)/(\omega-x)$ with $\omega \in (-\infty,0)$. Then, from \eqref{addf} we use the subordination property \cite[Chapter~22]{Speicher} as ${\rm M}^N_{\matr \Lambda -\matr J}(\omega)\simeq {\rm M}^N_{\matr \Lambda}(\omega+{\rm R}^N_{\matr J}(-{\rm M}^N_{\matr \Lambda -\matr J}(\omega)))$ and take the limit $\omega \to 0$. Doing so yields $V\simeq{\rm R}(1-q)$ where we note that ${\rm R}(x)=\lim_{N\to \infty} {\rm R}^N_{\matr J}(x)$. Notice also that
\begin{equation}
\frac{\ln \det (\matr \Lambda-V\matr {\bf I})}{\partial \Lambda_{ii}}=\frac{1}{\Lambda_{ii}-V}-N(1-q)\frac{\partial V}{\partial\Lambda_{ii}}.
\end{equation}
Hence, we finally obtain
\begin{align}
\frac{\partial \ln \det (\matr \Lambda-\matr J)}{\partial \Lambda_{ii}}&\simeq\frac{1}{\Lambda_{ii}-V}
\end{align}
which yields $V_i\simeq V = {\rm R}(1-q)$.

\begin{lemma}\label{Lemma1} 
	Let an $N\times N$ matrix $\matr X$ be positive definite. Let $Q=\frac{1}{N}{\rm tr}(\matr X^{-1}) $. Then,
	\begin{equation}
	\frac{1}{N}\ln \det (\matr X)=-(1+\ln Q)+ \int_{0}^{Q}{\rm d}\omega\; {\rm R}^N_{\matr X}(-\omega). \label{nice1}
	\end{equation} 
	\begin{proof}
		Note that 
		\begin{equation}
		\ln \det (\matr X)=\lim_{\epsilon\to \infty} \ln \det (\epsilon^{-1} \matr {\bf I}+\matr X).
		\end{equation}
		For convenience let $\eta (\epsilon)\triangleq\frac{1}{N} {\rm tr}((\matr {\bf I}+\epsilon\matr X)^{-1})$ for $\epsilon>0$. Since $\matr X$ is positive definite we can write \cite{tulino}
		\begin{equation}
		{\rm R}^N_{\matr X}(-\epsilon \eta(\epsilon))=\frac{1- \eta(\epsilon)}{\epsilon \eta(\epsilon)}.
		\end{equation}
		Applying the substitution $\omega\triangleq t \eta (t)$ to the following integral we have 
		\begin{align}
		\int_{0}^{Q(\epsilon)}{\rm d}\omega\;{\rm R}^N_{\matr X}(-\omega)&= \int_{0}^{\epsilon}{\rm d}t\;\frac{\eta(t)+t\eta'(t)}{t \eta(t)}[1-\eta(t)] \\
		&= \int_0^{\epsilon}{\rm d}t\;\frac{\eta'(t)}{\eta(t)}(1-\eta(t))+\int_{0}^{\epsilon}{\rm d}t\;\frac{1-\eta(t)}{t} \\
		&= \ln \eta(\epsilon)+ 1-\eta(\epsilon)+\int_{0}^{\epsilon}{\rm d}t\;\frac{1-\eta(t)}{t} \\
		&= \ln \eta(\epsilon)+ 1-\eta(\epsilon)+\frac{1}{N}\ln \det (\matr {\bf I}+\epsilon \matr X) \\
		&= \ln Q(\epsilon)+1-\eta(\epsilon)+\frac{1}{N}\ln \det (\epsilon^{-1}\matr {\bf I}+\matr X)
		\end{align}
		with $Q(\epsilon)\triangleq\epsilon \eta (\epsilon)$. In other words we have
		\begin{equation}
		\frac{1}{N}\ln \det (\epsilon^{-1}\matr {\bf I}+\matr X)= (\eta(\epsilon)-1-\ln Q(\epsilon))+\int_{0}^{Q(\epsilon)}{\rm d}\omega{\rm R}^N_{\matr X}(-\omega). \label{nice2}
		\end{equation}
		Taking the limit $\epsilon \to \infty$ we complete the proof.
	\end{proof} 
\end{lemma}

\section{Derivation of DFT Results} \label{devDFT}
For the sake of notational compactness let 
\begin{align}
{\rm g}(\left\{\matr m(\tau), \matr \gamma(\tau)\right\}_{\tau=0}^t)\triangleq \delta (\matr m(t)- f \left(\{\matr\gamma(\tau), \matr m(\tau) \}_{\tau=0}^{t-1}\right)).
\end{align} 
By the Fourier representation of the Dirac function we write 
\begin{align}
Z(\{\matr l(t)\})=\int \prod_{t=0}^{T-1}{\rm d}\matr m(t){\rm d}\matr \gamma(t){\rm d}\matr {\hat\gamma}(t)\;{\rm g} ( \left\{\matr m(\tau), \matr \gamma(\tau)\right\}_{\tau=0}^{t}) e^{i\matr {\hat\gamma}(t)^\dagger(\matr{\gamma}(t)-\matr h-\matr J \matr m(t))}
e^{i\matr\gamma(t)^\dagger \matr l(t)}.
\end{align}
The derivation is separated into two parts: i) disorder average; ii) the saddle point method. 
\subsection{disorder average}
For convenience let us introduce $N \times T$ matrices $\matr X$ and $\matr {\hat X}$ with $X_{nt}=\frac{m_n(t+1)}{\sqrt{N}}$ and $\hat X_{nt}= \frac{\hat \gamma_n(t+1)}{i\sqrt{N}}$.
We need to evaluate
\begin{equation}
 \left<e^{-\frac{i}{2}\sum_{t}\left\{\matr {\hat\gamma}(t)^\dagger\matr J \matr m(t)+\matr {m}(t)^\dagger\matr J \matr {\hat\gamma}(t)\right\}}\right>_{\matr J}\simeq e^{\frac{N}{2} \sum_{n\geq 1} \frac{c_n}{n} {\rm tr}(\matr Q^n)} \label{D1}
\end{equation}
with $\matr Q=\matr{\hat X} \matr {X}^\dagger+\matr X\matr {\hat X}^\dagger$. Here \eqref{D1} follows directly from \eqref{G1}-\eqref{G3}.  We will evaluate
\begin{equation}
{\rm tr}(\matr Q^n)={\rm tr} \left((\matr {\hat X} \matr {X}^\dagger+\matr X\matr {\hat X}^\dagger)^n\right). \label{key}
\end{equation}
in terms of the matrices \eqref{key}
\begin{align}
\mathcal {G}\triangleq\matr{X}^\dagger \matr {\hat X} \\
\mathcal{C}\triangleq\matr{X}^\dagger \matr {X} \\
\mathcal {\tilde{C}}\triangleq\matr{\hat X}^\dagger \matr {\hat X} 
\end{align}
Then by using cyclic invariance of the trace we obtain the expression
\begin{equation}
{\rm tr}(\matr Q^n)= 2{\rm tr}(\mathcal G^n)+ n{\rm tr} \sum_{k=0}^{n-2} \left\{\mathcal G^k \mathcal C (\mathcal G^\dagger)^{n-2-k} \mathcal {\tilde C}\right\}+ {\rm I}(\mathcal G, \mathcal C,\mathcal {\tilde C}) \ , \label{key1}
\end{equation}
where the function ${\rm I}$ satisfies
\begin{equation}
\left.\frac{\partial {\rm I}(\mathcal G, \mathcal C, \mathcal {\tilde C})}{\partial\mathcal {\tilde C}}\right\vert_{\mathcal {\tilde C}=0}=0.
\end{equation}
This means that ${\rm I}$ contains more than one factor $\mathcal {\tilde C}$ and will thus--at the saddle--point 
value--$\mathcal {\tilde C}=0$ not contribute to saddle--point equations.

\subsection{The saddle point calculation}
We write  as 
\begin{align}
&\left< Z(\{\matr l(t)\})\right>_{\matr J}\simeq \int  {\rm d} \mathcal G {\rm d}\mathcal C  {\rm d} \mathcal {\tilde C}\;e^{\frac{N}{2} \sum_{n\geq 1} \frac{c_n}{n}\left(2{\rm tr}(\mathcal G^n)+ n{\rm tr} \sum_{k=0}^{n-2} \left\{\mathcal G^k \mathcal C (\mathcal G^\dagger)^{n-2-k} \mathcal {\tilde C}\right\}+ {\rm I}(\mathcal G, \mathcal C,\mathcal {\tilde C})\right)} \nonumber  \\
& \int \prod_{t=0}^{T-1}\left\{{\rm d}\matr m(t){\rm d}\matr \gamma(t){\rm d}\matr {\hat\gamma}(t)\;{\rm g} (\left\{\matr m(\tau), \matr \gamma(\tau)\right\}_{\tau\leq t}) e^{i\matr {\hat\gamma}(t)^\dagger(\matr{\gamma}(t)-\matr h)}e^{i\matr\gamma(t)^\dagger \matr l(t)} \right\}\nonumber \\
& \prod_{t,s} \delta \left(iN\mathcal G(t,s)-\matr m(t)^\dagger\matr{\hat\gamma}(s)\right) \delta \left(N\mathcal C(t,s) -\matr m(t)^\dagger \matr{m}(s)\right)\delta \left(N\mathcal {\tilde C}(t,s) +\matr{\hat\gamma}(t)^\dagger\matr{\hat\gamma}(s)\right).\label{dirac}
\end{align} 
By the Fourier representation of Dirac function we write the last line of \eqref{dirac} as
\begin{align}
c\int{\rm d}\mathcal {\hat G} {\rm d}\mathcal {\hat C} {\rm d}\mathcal {\hat {\tilde C}}\;
e^{\sum_{(t,s)}i\mathcal {\hat G}(t,s)\left(iN\mathcal G(t,s) -\matr m(t)^\dagger \matr{\hat\gamma}(s)\right)+i\mathcal{\hat C}(t,s)\left(N\mathcal C(t,s) -\matr m(t)^\dagger\matr{m}(s)\right)-\mathcal{\hat{\tilde C}}(t,s)\left(N\mathcal{\tilde C}(t,s) +\matr{\hat\gamma}(t)^\dagger\matr{\hat\gamma}(s)\right)}. \label{partion}
\end{align}
Here $c$ is a constant irrelevant for the saddle point calculation. We define the auxiliary single-site partition function 
\begin{align}
\tilde Z_{n}(l_n,\mathcal {\hat G},\mathcal {\hat C},\mathcal {\hat {\tilde C}})\triangleq& \int \prod_{t=0}^{T-1}\left\{ {\rm d}m_n(t){\rm d}\gamma_n(t){\rm d}{\hat\gamma}_n(t)\;{\rm g}(\left\{m_n(\tau), \gamma_n(\tau)\right\}_{\tau\leq t}) 
e^{i\hat\gamma_n(t)({\gamma}_n(t)-h_n)}e^{i\gamma_n(t)l_n(t)}\right\} \nonumber \\
&\qquad e^{-\sum_{(t,s)}[i\mathcal{\hat G}(t,s)m_n(t){\hat\gamma}_n(s)+i\mathcal {\hat C}(t,s)m_n(t){m}_n(s)+\mathcal{\hat{\tilde C}}(t,s){\hat\gamma}_n(t) {\hat\gamma}_n(s)]}.
\end{align}
In this way we can write \eqref{dirac} as
\begin{align}
\left< Z(\{\matr l(t)\})\right>_{\matr J}\simeq &\int {\rm d} \mathcal G {\rm d}\mathcal {\hat G} {\rm d} \mathcal C {\rm d}\mathcal {\hat C} {\rm d} \mathcal {\tilde C} {\rm d}\mathcal{\hat {\tilde C}}\;e^{\frac{N}{2} \sum_{n\geq 1} \frac{c_n}{n} \left(2{\rm tr}(\mathcal G^n)+ n{\rm tr} \sum_{k=0}^{n-2} \left\{\mathcal G^k \mathcal C (\mathcal G^\dagger)^{n-2-k} \mathcal {\tilde C}\right\}+ {\rm I}(\mathcal G, \mathcal C,\mathcal {\tilde C})\right)} \nonumber \\
 &\qquad  e^{N\sum_{(t,s)}[-\mathcal{\hat G}(t,s) \mathcal G_{t,s}+i\mathcal{\hat C}(t,s)\mathcal C(t,s) -\mathcal{\hat{\tilde C}}(t,s) \mathcal{\tilde C}(t,s)]+\sum_{n} \log\tilde Z_{n}(l_n, \mathcal {\hat G},\mathcal {\hat C }, \mathcal {\hat {\tilde C}})}.
\end{align}
In the large $N$ limit we can perform the integrations over $\mathcal G, \mathcal {\hat G},\mathcal C, \mathcal {\hat C},\mathcal {\tilde C}, \mathcal {\hat {\tilde C}}$ with the saddle point methods. Doing so yields: 
\begin{align}
\mathcal G(t,s)&=-\frac{i}{N}\sum_n\left<m_n(t)\hat\gamma_n(s) \right>_{\tilde Z_n}\\
\mathcal C(t,s)&=\frac{1}{N}\sum_n\left<m_n(t)m_n(s) \right>_{\tilde Z_n}\\
\mathcal{\tilde C}(t,s)&=-\frac{1}{N}\sum_n\left<\hat\gamma_n(t)\hat\gamma_n(s) \right>_{\tilde Z_n}. \label{unpsy}
\end{align}
with $\langle~\rangle_{\tilde Z_n}$ denoting the average with respect to the single-site partition function. Here the quantity \eqref{unpsy} has only the trivial solution that $\mathcal{\tilde C}_{t,s}=0$. Other solutions may violate the normalization $ Z(\{\matr l(t)=0\})=1$. 
Furthermore this solution leads to $\mathcal {\hat C}=\matr 0$. 
By invoking \eqref{key1} we have
\begin{align}
\mathcal{\hat G}&= {\rm R}(\mathcal G)\\
\mathcal{\hat{\tilde C}}&=\frac{1}{2}\sum_{n=1}^\infty c_n\sum_{k=0}^{n-2}\mathcal G^k \mathcal C(\mathcal G^\dagger)^{n-2-k}.
\end{align}
Thus, for large $N$ we get the factorization of the generating function
\begin{align}
\left< Z(\{\matr l(t)\})\right>_{\matr J}\simeq & \prod_{n=1}^{N}\int \prod_{t=0}^{T-1}\left\{{\rm d}m_n(t){\rm d}\gamma_n(t)\;{\rm g}(\left\{m_n(\tau), \gamma_n(\tau)\right\}_{\tau\leq t})e^{i\gamma_n(t)l_n(t)}  
\right. \nonumber \\
&\qquad\qquad  \qquad\left. e^{i\hat\gamma_n(t)({\gamma}_n(t)-h_n-\sum_{s}\mathcal{\hat G}(t,s)m_n(s))} e^{-\sum_{s}\mathcal{\hat{\tilde C}}(t,s){\hat\gamma}_n(t){\hat\gamma}_n(s)}\right\}. \label{Avp}
\end{align}
We linearize the quadratic terms in $\hat\gamma_n(t)$ by introducing auxiliary Gaussian random fields $\phi_n(t)$ which are iid for each $n$ with zero mean and covariance $\mathcal C_\phi(t,s) \triangleq 2 \mathcal{\hat{\tilde C}}(t,s) \triangleq \langle{\phi}_n(t){\phi}_n(s)\rangle$. In this way we can write
\begin{equation}
e^{-\sum_{t,s}{\mathcal{\hat{\tilde C}}(t,s){\hat\gamma}_n(t){\hat\gamma}_n(s)}}= \left< e^{-i \sum_{t} \phi_n(t){\hat\gamma}_n(t)}\right>_{\phi_n} \ . 
\end{equation}
Doing so leads \eqref{Avp} to
\begin{align}
\left< Z(\{\matr l(t)\})\right>_{\matr J}\simeq \prod_{n=1}^N\int{\rm d}\mathcal N(\{\phi_n (t)\}; 0,\mathcal C_{\phi}) \prod_{t=0}^{T-1}\left\{{\rm d}m_n(t){\rm d}\gamma_n(t)\; {\delta}( m_n(t)-f_t\left\{m_n(\tau), \gamma_n(\tau)\right\}_{\tau=0}^{t-1})
\right. \nonumber \\
\left.  \delta\left(\gamma_n(t)-h_n-\sum_{s<t}\mathcal{\hat G}(t,s) m_n(s)-\phi_n(t)\right)e^{i\gamma_n(t){l}_n(t)} 
\right\}.
\end{align}
Finally, notice that $-i\langle m_n(t)\hat\gamma_n(s)\rangle_{\phi_n}= \langle \frac{\partial m_n(t)}{\partial \phi_n(s)}\rangle_{\phi_n}$. Thus, $\mathcal G$ equals the response--function \eqref{response}.  This completes the derivation.

\section{Derivation of {equation} \eqref{ssmcov}}\label{proofssmcov}
First note that from \eqref{psi} and \eqref{CoeffA} we have
\begin{align}
\mathcal C_{\psi}(t-1,t'-1)=\frac{(\mathcal G \mathcal C_{\phi} \mathcal G^\dagger)(t,t')}{(1-q(t))(1-q(t'))} \label{cov1} 
\end{align}
where $\mathcal C_{\phi}$ is defined as in \eqref{cov}. For sake of compactness we let $f(x)={\rm R}^{-1}(x)$. By elementary combinatorics 
and using 
$\mathcal{G}= {f}(\mathcal{\hat G})$
we can show that any power of the matrix $\mathcal G$ can be written as 
\begin{align}
\mathcal{G}^k = \sum_{n=1}^{\infty} {\rm Co}_{x^n} (f(x)^k)\mathcal {\hat G}^n. 
\end{align}
where for any power series $f(x) = \sum_n a_n x^n$ we represent the coefficient  
via the definition $a_k \triangleq {\rm Co}_{x^k} (f(x))$. 
This means that we have
\begin{align} 
{\mathcal G}^k(t, \tau)={\rm Co}_{x^{t-\tau}} ({f}(x)^k)\prod_{s=\tau}^{t-1}\mathcal {\hat G}(s+1,s).
\end{align}
Hence we also get
\begin{align}
(\mathcal {G}^{k+1}\mathcal {C}(\mathcal {G}^\dagger)^{n-1-k})(t-1,t'-1)= &\sum_{l<t, m<t'} 
\left(\prod_{s=l}^{t-1}\mathcal {\hat G}(s+1,s)\right)\left(\prod_{s=m}^{t'-1} \mathcal {\hat G}(s+1,s)\right)\nonumber \\
& \times \mathcal C(l,m){\rm Co}_{x^{t-l}y^{t'-m}} ({f}(x)^{k+1}{f}(y)^{n-1-k}).
\end{align}
where we have extended the definitions of coefficients ${\rm Co}_{x^ny^k} [f(x,y)]\triangleq a_n b_k$ to double power series
$f(x,y)=\sum_{n,k\geq 0}a_nb_k x^ny^k$.
Summing up the geometric series, we have 
\begin{equation}
\sum_{k=0}^{n-2}{\rm Co}_{x^{t-l}y^{t'-m}} ({f}(x)^{k+1}{f}(y)^{n-1-k})={\rm Co}_{x^{t-l}y^{t'-m}} \left(\frac{{f}(y)^{n-1}-{f}(x)^{n-1} }{[{f}(y)-{f}(x)]/[{f}(x){f}(y)]}\right)
\end{equation}
and
\begin{equation}
\sum_{n=1}^{\infty}c_n {\rm Co}_{x^{t-l}y^{t'-m}} \left(\frac{{f}(y)^{n-1}-{f}(x)^{n-1}}{[{f}(y)-{f}(x)]/[{f}(x){f}(y)]}\right)={\rm Co}_{x^{t-l}y^{t'-m}}\left(\frac{y-x}{[{f}(y)-{f}(x)]/[{f}(x){f}(y)]}\right).
\end{equation}
Putting everything together completes the proof.

\section{Derivation of equation \eqref{comega}}\label{ATdetails}
For the sake of compactness, without loss of generality, we may set $h_i=h$. 
Using the representation of the Gaussian density in terms of the characteristic function we have the expansion
for $t\neq t'$
\begin{align}
&\left\langle \tanh(u(t) + h) \tanh(u(t')+ h) \right\rangle_{u}  \simeq \nonumber  \\
 &\int {\rm d} u_1 {\rm d}u_2 {\rm d}k_1 {\rm d}k_2\; \exp\left[i (k_1 u_1 + k_2 u_2) - \frac{1}{2} \{s_0 + \epsilon (s(t,t) +1)\}(k_1^2 + k_2^2) -
\{s_0 + \epsilon s(t,t')\}  k_1 k_2
\right]  \times \nonumber  \\
& \times \tanh(u_1+ h_i) \tanh(u_2 + h_i) \nonumber \\
&\simeq  \; q - \frac{\epsilon}{2} \int {\rm d}u_1 {\rm d}u_2 {\rm d}k_1 {\rm d}k_2\; \exp\left[i (k_1 u_1 + k_2 u_2) - \frac{s_0}{2} (k_1^2 + k_2^2 + 2 k_1 k_2) 
\right]  \times \nonumber \\
 &\times \tanh(u_1+ h) \tanh(u_2 + h) \left\{(s(t,t) +1)(k_1^2 + k_2^2) + 2 s(t,t')  k_1 k_2\right\} \nonumber \\
&=  \; q + \epsilon (s(t,t) + 1) \left\langle\tanh(u + h) \frac{\partial^2 \tanh(u + h)}{\partial u^2} \right\rangle
+ \epsilon s(t_1,t_2) \left\langle\left(\frac{\partial \tanh(u + h)}{\partial u}\right)^2\right\rangle \ . 
\end{align}
The last line is obtained by representing $k_1 k_2$ etc by derivatives with respect to $u_1$ and $u_2$.
Repeating a similar equation for $t=t'$, we get
\begin{align}
&\left\langle \tanh^2(u(t) + h) \right\rangle_{u}  = 
q +
\frac{\epsilon}{2} (s(t,t) + 1) \left\langle\frac{\partial^2 \tanh^2(u + h)}{\partial u^2} \right\rangle = \nonumber \\
& = q +
\epsilon (s(t,t) + 1)\left\langle\tanh(u + h) \frac{\partial^2 \tanh(u + h)}{\partial u^2}  + \left(\frac{\partial \tanh(u + h)}{\partial u}\right)^2 \right\rangle.
\end{align}
Both expansions can be represented in the single equation
\begin{align}
\frac{{\rm d}\left\langle \tanh(u(t) + h) \tanh(u(t')+ h) \right\rangle_{u} }{{\rm d}\epsilon} &=(s(t,t) + 1) \left\langle  \tanh(u + h) \frac{\partial^2 \tanh(u + h)}{\partial u^2} \right\rangle
\nonumber \\
& + (s(t,t')+ \delta_{t,t'} )\left\langle  \left(\frac{\partial \tanh(u + h)}{\partial u}\right)^2   \right\rangle .
\end{align}
Note, that only the second term contributes to the dynamic part of the fluctuations. Hence, by taking the Fourier transform 
and noting that $\frac{\partial \tanh(u + h)}{\partial u} = 1- \tanh^2(u+h)$ the result is obtained.

\bibliographystyle{iopart-num}
\bibliography{iopart-num}

\end{document}